\documentclass[twocolumn,trackchanges]{aastex631}

\usepackage{amsmath}
\usepackage{lipsum}
\usepackage{graphics}
\usepackage{CJK}

\received{January 19, 2025}
\revised{March 4, 2025}

\usepackage{graphicx}

\begin{document}

\title{The $M_{*}-M_{\rm BH}$ Relation Evolution from z $\sim$ 6 to the Present Epoch }

\correspondingauthor{Yang Sun}
\email{sunyang@arizona.edu}

\author[0000-0001-6561-9443]{Yang Sun}
\affiliation{Steward Observatory, University of Arizona,
933 North Cherry Avenue, Tucson, AZ 85719, USA}

\author[0000-0003-2303-6519]{George H. Rieke}
\affiliation{Steward Observatory, University of Arizona,
933 North Cherry Avenue, Tucson, AZ 85719, USA}

\author[0000-0002-6221-1829]{Jianwei Lyu (\begin{CJK}{UTF8}{gbsn}吕建伟\end{CJK})}
\affiliation{Steward Observatory, University of Arizona,
933 North Cherry Avenue, Tucson, AZ 85719, USA}

\author[0000-0002-9720-3255]{Meredith A. Stone}
\affiliation{Steward Observatory, University of Arizona,
933 North Cherry Avenue, Tucson, AZ 85719, USA}

\author[0000-0001-7673-2257]{Zhiyuan Ji}
\affiliation{Steward Observatory, University of Arizona,
933 North Cherry Avenue, Tucson, AZ 85719, USA}

\author[0000-0002-5104-8245]{Pierluigi Rinaldi}
\affiliation{Steward Observatory, University of Arizona, 933 North Cherry Avenue, Tucson, AZ 85719, USA}

\author[0000-0001-9262-9997]{Christopher N. A. Willmer}
\affiliation{Steward Observatory, University of Arizona, 933 North Cherry Avenue, Tucson, AZ 85719, USA}

\author[0000-0003-3307-7525]{Yongda Zhu}
\affiliation{Steward Observatory, University of Arizona,
933 North Cherry Avenue, Tucson, AZ 85719, USA}



\begin{abstract}
The ratio between the stellar mass of a galaxy, $M_{*}$,  and that of its central supermassive black hole (SMBH), $M_\bullet$, the ``Magorrian'' relationship, traces their coevolution. JWST observations have suggested significant evolution in $M_\bullet/M_{*}$ relative to local scaling relationships both in low-mass galaxies and in quasars at z $\ge$ 4. We test this possibility by (1) determining  the preferred $M_\bullet/M_{*}$ scaling relation among those proposed locally; and (2) providing uniform host galaxy stellar mass estimates. These steps reduce the prominence of the reported evolution. We then apply Monte Carlo simulations to account for observational biases. We still find a significant increase over the local scaling relation in $M_\bullet/M_{*}$  for z $\ge$ 4 SMBHs in very low-mass galaxies ($\log(M_*/M_{\odot})<10$). However, similarly high values of $M_\bullet/M_{*}$ are also found in low mass galaxies at $z \sim$ 0.5 to 3 that may be common at cosmic noon. Nonetheless, galaxies with similar behavior are rare locally and not accounted for in the local scaling relations. In contrast, z $\sim$ 6 quasars can have $M_\bullet/M_{*}$ well above the local relation value, but they can be explained as extreme cases still within the scaling relation for their  higher mass host galaxies. Black holes in some of them and in the low-mass systems may be undergoing very high accretion episodes that result in high $M_\bullet/M_{*}$ but that will be followed by quiescent periods when growth of the host drives the systems toward more typical  $M_\bullet/M_{*}$ values.


\end{abstract}



\section{Introduction} \label{sec:intro}

The discovery of quasars led quickly to the suggestion that they derived their power from accretion onto very compact, massive objects in the nuclei of galaxies \citep{Hoyle1963, Salpeter1964}. This idea was confirmed with the finding through dynamical measurements that super massive black holes (SMBHs) lurk in the nuclei of many massive galaxies \citep{Kormendy1995}. The correlation between SMBH masses ($M_\bullet$) and the masses of their host galaxies ($M_{*}$), especially the masses of the galaxy bulge component ($M_{b}$), was discovered by \citet{Magorrian1998}. It has been confirmed and expanded by many works \citep[e.g.,][]{Ferrarese2000,Gebhardt2000,Haring2004,KH2013}, suggesting that the central black holes (BHs) co-evolve with their hosts. However, the physical mechanism underlying galaxy-BH co-evolution is not well understood. There are two possibilities: (1) that feedback (outflow from the black hole) regulates the host galaxy properties \citep[e.g.,][]{Springel2005,Hopkins2008}; or (2) that the galaxy mass assembly (e.g., mergers) links the growth of central supermassive black holes with host galaxies, without a physical coupling between them \citep{Peng2007,Jahnke2011}. Understanding how this relation evolves with time (or redshift) is central to extragalactic astronomy, and is now challenged by the discovery with JWST of more very high redshift AGNs than anticipated \citep[e.g.,][]{Ignas2023, Larson2023,
 Scholtz2023, Matthee2024, Treiber2024, Grazian2024,Taylor2024}, possibly with SMBHs that are overmassive compared with their hosts \citep{Harikane2023, Pacucci2023, Kokorev2023, 
 Furtak2024, Natarajan2024}. 

Testing how ($M_\bullet$/$M_{*}$) changes with redshift requires us to measure the relation at different redshift ranges accurately. In the local Universe (z $<$ 1), $M_{\rm BH}$ measurements can be done accurately by benchmarking among several methods, such as reverberation mapping (RM), single-epoch (SE) virial BH masses through broad spectral lines, and dynamical BH masses estimated from the nuclear stellar velocity dispersion (see \citealt{Shen2013} for a detailed review). $M_{b}$ can also be well constrained by photometric bulge-disk decomposition. Therefore, the $M_{\bullet}$-$M_{b}$ relation is well determined at z $<$ 1 (\citealp{JLi2023}; see \citealt{KH2013} for a detailed review). 

However, at higher redshift, black hole mass can usually only be estimated with the SE virial method for broad-line (BL) AGNs. Also, due to the lack of spatial resolution and sensitivity, $M_{b}$ is hard to constrain. Instead, the total stellar mass ($M_{*}$), measured by stellar population synthesis from SED modeling or assuming a galaxy mass-to-light ratio from the observed photometry, is commonly used to trace the $M_\bullet$-$M_{*}$ relation. Within these limitations, the relation has been thoroughly measured up to z $\sim$ 4,  and appears to show little redshift evolution (see discussion in e.g., \citealt[][and references therein]{Sun2025, li2025}).

 \emph{JWST} offers new opportunities to study AGNs down to faint luminosities and to much higher redshifts. Many works have found faint BL AGNs at $z>4$ using \emph{JWST}/NIRSpec or NIRCam/WFSS spectroscopic data at $\lambda < $ 5$\mathrm{\mu m}$ \citep[e.g.,][]{Carnall2023,Harikane2023,Kocevski2023,Larson2023,Maiolino2023, Ubler2023, Taylor2024}. Host galaxies around  $z\sim6$ quasars are also starting to be found and measured with deep Near-Infrared (NIR) imaging \citep{Ding2023,Yue2024,Stone2024,Marshall2024, Onoue2024}.

 \emph{JWST} observations have suggested that, beyond $z$ $\sim$ 4, the typical $M_\bullet$/$M_{*}$ may be well above the average value for $1 < z < 4$. \citet{Pacucci2023} and \citet{Harikane2023} report systematic surveys indicating that AGNs at $4 < z < 7$ have substantially over-massive black holes, and similar behavior is reported for the high redshift quasars now found in substantial numbers \citep{Fan2023}, see \citet{Marshall2023, Ding2023,
 Yue2024, Stone2024, Marshall2024,Onoue2024}, as well for isolated examples at even higher redshift \citep[e.g.,][]{Kokorev2023, Furtak2024, Natarajan2024}. These results have significant implications for our understanding of how black holes and galaxies co-evolve \citep[e.g.,][]{Hopkins2008, Habouzit2021}. However, given the long controversy regarding $M_\bullet$/$M_{*}$ at $z < 2$, this conclusion is not definitive, and has been questioned by some  \citep{Li2024}.

 In this paper, we test the behavior of $M_\bullet$/$M_{*}$ at $z > 4$ in detail. We first describe the sample of high redshift AGNs we will analyze (Section 2).  In section 3, we evaluate the local relation that is the foundation for any evidence for changes at high redshift. The scaling relation due to  \citet[][hereafter GSH20]{Greene2020} is strongly preferred and we use it to examine the behavior of 
$M_\bullet$/$M_{*}$ at $z > 4$. In Section 4, we unify the galaxy and black hole mass estimates at z $>$ 4 and test the revised trend of apparent $M_\bullet/M_{*}$ for biases. To do so, we use an expansion of the Monte Carlo method proposed by \citet{Li2021} to determine selection biases, such as that identified by \citet{Lauer2007}. We discuss other effects that might result in erroneous results in Section 5.   Section 6 summarizes our results. 

Throughout this paper, we assume a standard
$\Lambda$CDM universe with cosmological parameters H$_0$ =
70 km s$^{-1}$ Mpc$^{-1}$, $\Omega_\Lambda$ = 0.7, and $\Omega_m$ = 0.3.

\section{High redshift samples}
\label{sec:comparison_sample}

To test the redshift evolution of the $M_{\bullet}$-$M_{*}$ relation over $4<z<7$, we will discuss three distinct samples: (1) AGNs with modest mass black holes in low-mass galaxies; for brevity, we will describe this sample as {\bf ``Seyfert galaxies''} or just {\bf ``Seyferts''}; (2) Little Red Dots ({\bf LRDs}); and (3) high luminosity AGNs in quasars, with moderately massive host galaxies; to be termed {\bf ``quasars''} in the following.  We will also include previously published studies at lower redshift. 

\subsection{High-z Seyfert Galaxies}

This sample is identical to the one analyzed by \citet{Pacucci2023}, but we will carry out a new analysis. It consists of 17 moderate-luminosity AGNs: eight from \citet{Harikane2023} and nine  from \citet{Maiolino2023}\footnote{We exclude GLASS{\_}160133 which does not have proper imaging-based stellar mass measurement and CEERS{\_}01236, which has dual nuclei, from the \citet{Harikane2023} sample and three dual AGN candidates from the \citet{Maiolino2023} sample.} at $4.4<z<7$. These AGNs have broad H$\alpha$ or H$\beta$ lines,  and BH masses in the range of  $6.5<\log(M_{\bullet}/M_{\odot})<8$. Their  bolometric luminosities are $44.1<\log (L_{bol}/\mathrm{erg\,s^{-1}})< 46.$ They were discovered in deep JWST surveys, including the ERO programs  \citep{Pontoppidan2022}  CEERS \citep{Finkelstein2023} and GLASS \citep{Treu2022}, and in the JADES GTO program \citep{Eisenstein2023}.

For these AGNs, we obtained the BH masses from \citet[][Table 2]{Harikane2023} and \citet[][Table 3]{Maiolino2023}, and galaxy masses from \citet[][Table 3]{Harikane2023} and \citet[][Table 3]{Maiolino2023}. However, there are small inconsistencies in the methods used to estimate these parameters from paper to paper. We will update them to a common basis as described in Section~\ref{unification}.

\subsection{Little Red Dots}

Since their discovery \citep{Labbe2023}, there has been intensive study of very compact sources at z$>$4 discovered by JWST and characterized by  extremely red color at the longer wavelengths and a much bluer color at short ones - often described as a V-shaped 
SED.  \citet{Matthee2024} created the term ``LRD" to describe the red and compact appearance of these systems.  They have been found in large numbers \citep[e.g.,][]{Akins2024}, but their  nature is under debate \citep[e.g.,][]{Killi2024, Kokorev2023, Kokorev2024, Hainline2025}. The observations of photometrically-selected LRDs have shown that a significant fraction do exhibit broad Balmer lines indicative of AGNs \citep[e.g., ][]{Greene2024,Kocevski2024,Rinaldi2024}. However, photometry at wavelengths longer than 5 $\mu$m has shown that many LRDs have SEDs that turn over past 7 $\mu$m, as expected if their SEDs are dominated by stars \citep{williams2024, Perez-Gonzalez2024}. That is they appear to have highly diverse and complicated origins \citep[][and references therein]{Perez-Gonzalez2024}.  Only those LRDs with broad Balmer lines, making them very likely to be AGN and enabling black hole mass estimation, qualify for discussion in the mass scaling relation.
However, even the BL LRDs, in general, do not combine the necessary attributes of significant numbers, unbiased selection, and homogeneous determinations of black hole and host galaxy masses for our study.

We have therefore focused on the work of \citet{Chen2024}. They  provided black hole masses and (mostly upper limits for) host galaxy masses for a sample of BL LRDs drawn from the Ultradeep NIRSpec and NIRCam ObserVations before the Epoch of Reionization (UNCOVER; \citealt{Bezanson2024}) program. They described how the sample was down-selected through color cuts, PSF fitting, and spectroscopy to the 12 AGNs that are the targets for their analysis. They therefore provided the necessary parameters for a sample drawn in a way that should be unbiased. However, all but one of their LRDs have only upper limits for host galaxy mass. We will compare these LRDs with the Seyferts. However, they did not provide a sufficiently large sample with stellar mass measurement to simulate independently the observational biases on the mass scaling relation.

\subsection{High-z quasars}
\label{quasarprops}

This sample includes 12 luminous quasars at $5<z<7.1$: two from \citet{Ding2023}, nine from \citet{Yue2024},  \citet{Stone2024}, and \citet{Marshall2024}, and one more from \citet{Onoue2024}, with bolometric luminosities $\log (L_{bol}/\mathrm{erg\,s^{-1}})> 46$. Their black hole masses are preferably measured from a broad H$\beta$ or H$\alpha$ line, or otherwise from MgII or CIV and are in the range of $8.3<\log(M_{\bullet}/M_{\odot})<10$. This sample is neither area-limited nor flux-limited; these quasars are selected from various surveys for specific high-redshift science objectives. 

Five of the galaxy mass estimates for the quasars at z $\sim$ 6 are upper limits. We have treated the upper limits as if they are actual mass measurements, which should be a conservative approach since, in general, one would assume that the masses are {\it smaller} than the upper limits on them. Further details about the quasar host galaxy masses are provided in Section~\ref{qsohostmasses}.

\section{The $M_\bullet - M_{*}$ relation}\label{sec:baseline}

The assumed baseline $M_\bullet - M_{*}$ relation is critical to judge whether there is evolution. In this section, we discuss first what is known about the relation locally and at modest redshift, and then explore the application of this  relation to AGNs up to z $\sim$ 7.

\subsection{Local Mass Scaling Relation for  galaxies with $\log(M_*/M_{\odot}) > 10$}\label{sec:scaling_relation_high_mass}

\begin{figure}
    \centering
    \includegraphics[width=1\hsize]
    {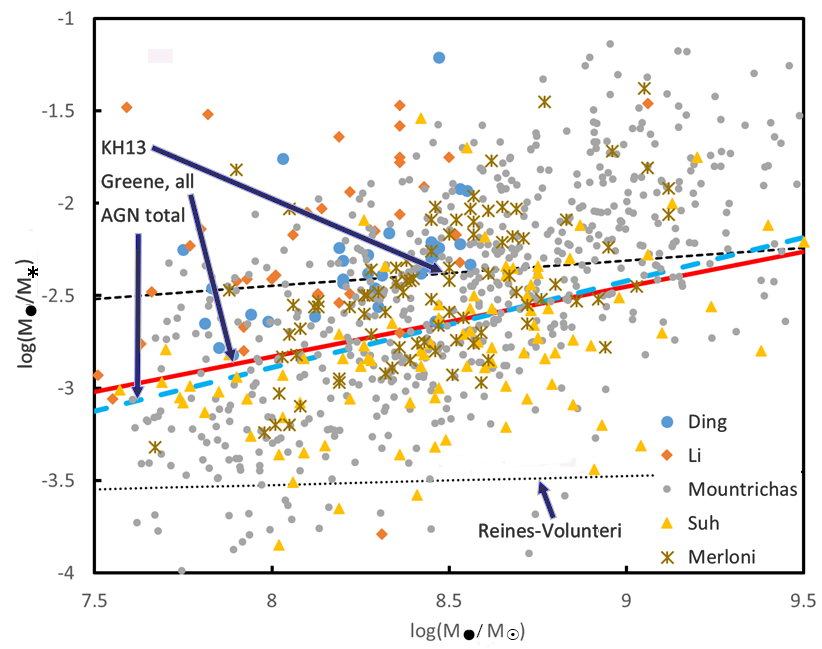}
       \caption{Dependence of $M_\bullet/M_{*}$ on $M_\bullet$ at z$\sim$2. Individual points are shown for the studies by \citet{Ding2020}, \citet{JLi2023}, \citet{Mountrichas2023}, \citet{Suh2020}, and \citet{Merloni2010}. They are compared with the scaling relations derived by \citet{Reines2015},  \citet{KH2013} for bulges, and  \citet{Greene2020}  for ``all, limits'' in heavy red and a fit to the total of five sets of individual measurements (AGN total). In this fit, we have used subsets of the \citet{Mountrichas2023} sample (of $\sim$ 600) so it does not dominate. This sample has a steeper slope than the other studies, leading to an apparent mis-match of the fitted slopes with the full set of points.}
       \label{mbh_mstars}
\end{figure}

The foundation for studies of  $M_\bullet/M_{*}$ is the local relation, against which the behavior at high redshift is compared to test for evolution. Our discussion of these relations is illustrated by Figure~\ref{mbh_mstars}.

Although the scaling relation between $M_\bullet$ and bulge mass has been shown to be tighter than that based on $M_{*}$ \citep{KH2013}, most studies at high redshift are unable to resolve bulges systematically and are based on total masses as a result. Both for this reason and because it extends to relatively low masses, the local relation determined by \citet{Reines2015} (hereafter RV15) has been widely used; it is based on an unbiased sample of local AGNs (z $<$ 0.055) and is in terms of the full mass of the host galaxies, rather than their bulge masses. The RV15 sample was dominated by relatively low-mass late-type AGN-host galaxies, for which the black hole mass lies largely between $10^6$ and $3 \times 10^7$ M$_\odot$ and the stellar mass between $10^{9.5}$ and $10^{11}$ M$_\odot$. For the ``uniformly selected'' AGN systems, they found a flat relationship, with $M_\bullet \propto M_{*}^{1.05 \pm 0.11}$.  

The RV15 relation is still used in many studies. However, it has largely been supplanted by the relation from \citet{Greene2020}, (hereafter GSH20).  The expanded study of GSH20 includes a large number of additional cases, including those with $\log(M_\bullet/M_{\odot}) \gtrsim 9$. The relations from GSH20 are quite different from those in RV15 (see Equations 1 \& 2, which drop the intrinsic scatter terms). This difference is partly because they included upper limits in deriving the slope of the relation \citep{Greene2020}, resulting in a steeper slope than found by RV15.
For example, their equation for ``all'' galaxies is:

\begin{equation}
\begin{split}
\log\left(\frac{M_{\bullet, a}}{M_\odot}\right) = \left(7.43 \pm 0.09\right) \\
+\left(1.61 \pm 0.12\right)
\times \log\left(\frac{M_{*}}{3 \times 10^{10} M_\odot}\right) 
\end{split}
\end{equation}
\noindent
and for late-type galaxies
\begin{equation}
\begin{split}
\log\left(\frac{M_{\bullet, l}}{M_\odot}\right) = \left(6.70 \pm 0.13\right) \\
+ \left(1.61 \pm 0.24\right)
\times \log\left(\frac{M_{*}}{3 \times 10^{10} M_\odot}\right) 
\end{split}
\end{equation}

\noindent
where $M_{\bullet, a}$ is $M_\bullet$ for the all, limits case and $M_{\bullet, l}$ is $M_\bullet$ for the late, limits case as defined in \citet{Greene2020}. Both of these relations indicate $M_\bullet \propto M_{*}^{1.6}$.

It is important to test the validity of the GSH20 relations, including for high redshift.  Fortunately, \citet{Suh2020} have done that by taking the RV15 local results for low masses and their results for higher masses at $z \sim$ 1 - 2. They fit the ensemble with Equation 3, which when reconfigured yields Equation 4 in a form directly comparable with the GSH20  relations: 

\begin{equation}
\begin{split}
\log\left(\frac{M_\bullet} {M_\odot}\right) = - \left(10.29 \pm 0.04\right) \\
+ \left(1.64 \pm 0.07\right)
\times \log\left(\frac{M_{*}}{M_\odot}\right) 
\end{split}
\end{equation}

\begin{equation}
\begin{split}
\log\left(\frac{M_\bullet} {M_\odot}\right) = \left(6.89 \pm 0.04\right) \\
+ (1.64 \pm 0.07) \times \log\left(\frac{M_{*}}{3 \times 10^{10} M_\odot}\right)
\end{split}
\end{equation}

\noindent
The agreement between Equations 1 \&   2 and 4 is remarkable (with Equation 4 falling between the other two), fully consistent with the view the $M_\bullet/M_{*}$ is not roughly a constant but that it grows with stellar mass roughly to the 1.6 power. 

Even using just the high redshift ($z>>0$) samples (i.e., not the RV15 sample for low masses), we can test the relation, although not fully confirm it because of the lack of systems with really low mass black holes measured at high z. The slope from GSH20 in the coordinates of Figure~\ref{mbh_mstars} is $0.38 \pm 0.06$.  The various samples shown in Figure~\ref{mbh_mstars} have a range of individual slopes; it is likely that there are systematic errors, possibly connected with the means to isolate the host galaxies and estimate their masses. Nonetheless, a linear regression fit\footnote{using subsamples of the Mountrichas sample so its larger numbers do not dominate} yields a slope of $0.47 \pm 0.07$, in agreement with the GSH20 relation within the errors. 
Both slopes are shown in Figure~\ref{mbh_mstars} (the one derived from the high redshift data labeled as ``AGN total''),  which emphasizes that they have little difference over the range of $M_\bullet$ of interest. 

As shown in Figure~\ref{mbh_mstars}, extrapolating the RV15 relation to the typical study beyond the local Universe with $\log(M_\bullet/M_{\odot}) \sim 8 - 9$ leads to an order of magnitude discrepancy with the GSH20 relation, an effect that could incorrectly indicate a significant cosmic evolution of $M_\bullet/M_{*}$.

Adoption of the GSH20 relation along with the large sample collected in  Figure~\ref{mbh_mstars} allows a re-determination of the scatter around the local relation. We find it to be 0.8 dex, as has been widely used previously. This value is, in fact, identical to that determined for the GSH20 scaling relation for local galaxies \citep{Greene2020}. 
The scatter shown in Figure ~\ref{mbh_mstars} arises from both the intrinsic scatter of the relation, and BH and stellar mass measurement uncertainties, but it is apparent that different black hole or stellar masses have different $M_{\bullet}/M_{*}$ ratios.

\begin{figure}
    \centering
   \epsscale{1.1} \plotone{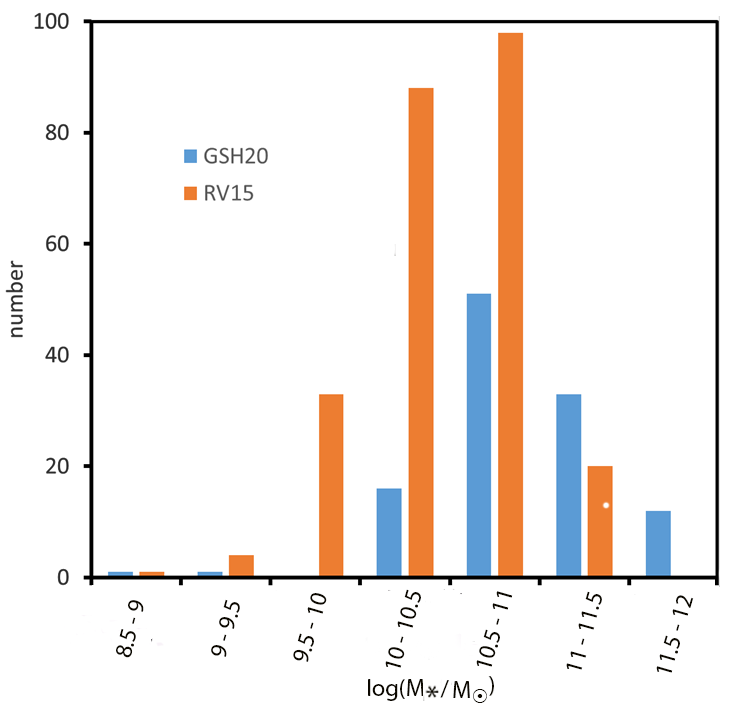}
       \caption{The distribution of host galaxy stellar masses of samples used to derive the local (within $\sim$ 100 Mpc) scaling relations of \citet{Reines2015, Greene2020}, RV15 and GSH20. Neither relation includes many  galaxies with $\log(M_{*}/M_\odot) < 9.5$\,  and the numbers are modest between 9.5 and 10. The GSH20 work adds significant weight to the high-mass end of the scaling relation. }
       \label{massdist}
\end{figure}

\subsection{Local Mass Scaling Relation at $log(M_{*}/M_{\odot}) < 10$} \label{sec:scaling_relation_low_mass}

Figure~\ref{massdist} shows the distribution of host galaxy stellar masses used in deriving the RV15 and GSH20 scaling relations. Clearly, we cannot expect these relations to be accurate in the low-mass galaxy ($\log(M_{*}/M_{\odot}) \le 9.5$) regime, where any application of them requires extrapolation with few observational constraints. In fact, at modest redshift ($z = 0.5 - 3)$, \citet{Mezcua2023, Mezcua2024} have found a population of AGNs in low mass galaxies that appear to be significantly overmassive relative to the local scaling relations. 

\subsection{Mass Scaling Relation up to z $\sim$ 7} \label{sec:scaling_relation_lz7}

The $z = 4 - 7$ Seyferts lie in host galaxies with $log(M_{*}/M_{\odot}) < 10$. For comparison with the previous studies by \citet{Harikane2023, Pacucci2023}, in our analysis of the behavior of these galaxies we will apply an extrapolation of the local scaling relation to low masses, as was done in those papers. This approach is also necessary because there is no well defined scaling relation that includes the examples from \citet{Mezcua2023, Mezcua2024}. However, the galaxies discussed by \citet{Mezcua2023, Mezcua2024} provide an alternate yardstick against which to measure evidence for evolution at high redshifts.

\begin{figure}
    \centering
   \epsscale{1.1} \plotone{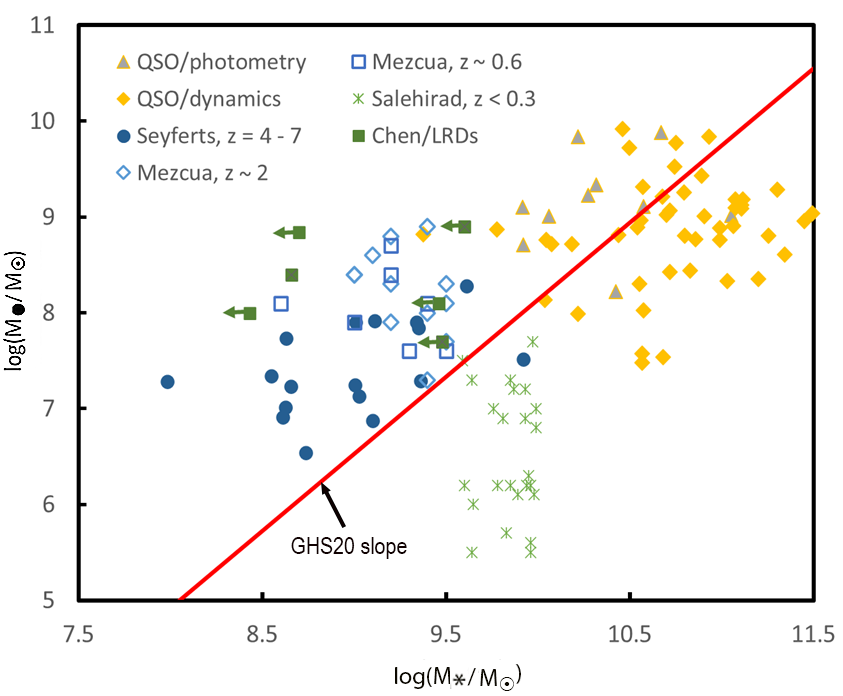}
   \caption{M$_\bullet$ vs M$_{*}$ for AGNs over a range of redshift. The data points are from (1) QSO/photometry:  \citet{Stone2024}: green triangles for values from infrared image deconvolution and QSO/dynamics: gold diamonds for masses from dynamics in the millimeter-wavelengths; (2) blue dots for z = 4 $-$ 7 Seyferts from \citet{Harikane2023, Pacucci2023}, after uniformization as described in Section~\ref{unification}; (3) open squares:    galaxies with 0.3 $<$ z $<$ 0.7 from \citet{Mezcua2023} and open diamonds: similar galaxies with 1 $<$ z $<$ 3 from \citet{Mezcua2024}; (4) brown squares: LRDs from \citet{Chen2024}; and (5) asterisks:  local galaxies from \citet{Salehirad2022}. The line is at the slope of the GSH20 scaling relation for $\log(M_{*}/M_{\odot}) > 10$, normalized to the high mass points. }
       \label{mvsm}
\end{figure}

Figure~\ref{mvsm} provides a summary of the situation. Given that the GSH20 relation is still valid for high-mass galaxies with $\log{M_{*}/M_{\odot}>10}$ at z $\sim$ 2 (see Section~\ref{sec:scaling_relation_high_mass}),
we normalized the GSH20 relation to the z $\sim$ 6 high mass sources from \citet{Stone2024}, as a way to avoid the observational biases in that sample
\footnote{It does not mean that the observational biases of the high-z high-mass quasars can be simply corrected by shifting the baseline relation upwards. Instead, the normalization is to illustrate that a single linear relation with the GSH20 slope fails to explain both the low-mass and high-mass galaxies at high redshift. See Section~\ref{sec:obs_bias_simu} for our comprehensive bias simulation.}. 
In this case, it is clear that, at high masses, the values roughly follow the slope of the GSH20 scaling relation after normalization. However, at lower masses, the values for the high redshift Seyferts depart substantially from this relation toward higher BH masses and the slope as a function of stellar mass is shallow.  The behavior at moderate redshifts of $\sim$ 0.3 - 3 is similar; it shows high ratios of $M_\bullet/M_{*}$ in low-mass galaxies, $8.6 \le \log(M_{*}/M_{\odot}) \le 9.5, $ \citep{Mezcua2023,  Mezcua2024}. \citet{Mezcua2024} believe that their results also apply ``to those samples of low-mass galaxies derived
using JWST data.'' Despite the similarity in mass, local dwarf galaxies behave quite differently, see \citet[][Figure 26]{KH2013} and \citet{Reines2022}, and the available studies \citep[e.g.,][]{Salehirad2022} indicate that their ratios of $M_\bullet/M_{*}$ are significantly lower than for the moderate redshift low-mass galaxies. For this reason, we distinguish between ``dwarf'' and ``low-mass'' galaxies in the following. 

The LRDs have low mass host galaxies ($\log(M_*/M_{\odot}) < 10$) and fall in the same region as the other AGNs in low-mass galaxies. There is even a hint that they may have more extreme (high) values of $M_\bullet/M_{*}$, a trend that needs confirmation with a larger sample. There are also indications that LRDs have different broad line characteristics than ``classical'' AGNs \citep{Taylor2024}. Given the possibility that LRDs behave systematically differently from ``classical''  AGNs, their $M_\bullet/M_{*}$ behavior may be unique. However, because of the lack of a good understanding of the distribution of LRD host galaxy masses, systematic analysis of this behavior is premature.

We will return to the issue of low-mass scaling relations in Section~\ref{modscale}, after we have explored the consequences of modeling with the extrapolated local relation.

\subsection{Summary of Scaling Relations}

We often talk about {\it the} Magorrian Relation. However, we have found that this is a major over-simplification. 
For galaxies with $\log(M_{*}/M_{\odot}) > 10$, the relation between $M_\bullet$ and $M_{*}$ derived by GSH20 is confirmed as the appropriate average behavior against which to measure possible evolution of the $M_\bullet/M_{*}$ relation, although there is a very large scatter of 0.8 dex around it.  Their relation for ``all'' systems also agrees very well with the normalization for massive systems at redshifts of $1 - 2$, $log(M_\bullet/M_{*}) \sim -2.5$ at $\log(M_\bullet/M_{\odot}) \sim 8.8$ (see \citealt{Sun2025}). 

However, the scaling relation is not well constrained for low-mass galaxies. For a given galaxy mass, there appears to be a significant change in the behavior of $M_\bullet/M_{*}$ already from the local neighborhood to z $\sim$ 1, as shown in Figure~\ref{mvsm}. 
The apparent discrepancy between the local behavior of dwarf galaxies and that of low mass ones at moderate redshift makes it difficult to derive a universal scaling relation for low masses. This contrasts with the situation for higher masses where, as we have just shown, the GSH20 relation derived locally is also applicable at $z \sim 1.5$.

\subsection{Consequences}

\begin{figure*}
\centering
    \includegraphics[width=0.7\textwidth]{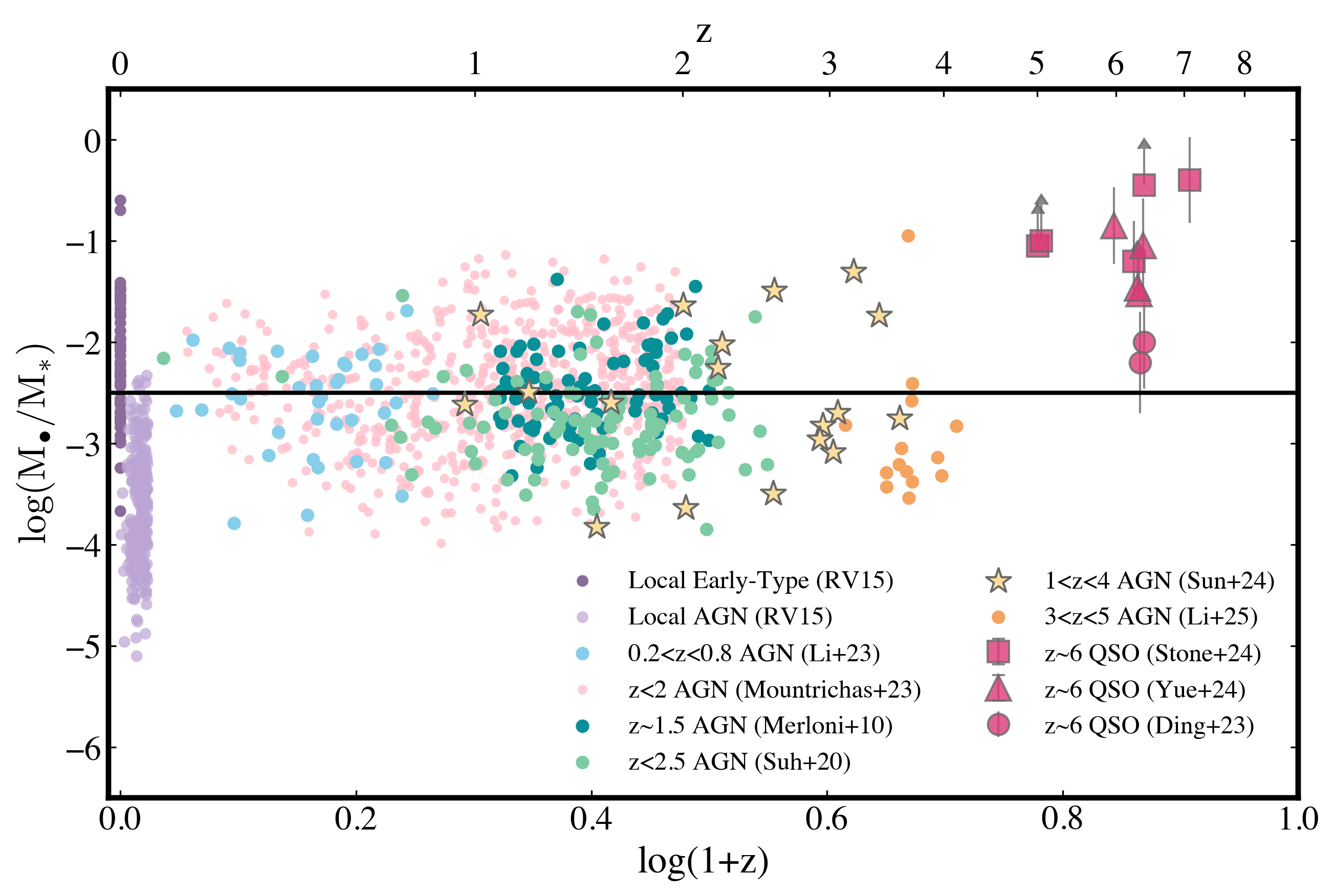}
\caption{$\log (M_{\bullet}/M_{*})$ as a function of redshift for massive host galaxies ($\log(M_*/M_\odot) \gtrsim 10$). 
The z$\sim$6 quasar samples from \citet{Ding2023}, \citet{Yue2024}, and \citet{Stone2024} are plotted by magenta circle, triangle, and square, respectively. For comparison, we illustrate several comparison samples from previous studies for $M_{\bullet}/M_*$, whose redshift ranges from z$\sim$0 to 4 \citep{Reines2015,Merloni2010,Suh2020,JLi2023,Mountrichas2023,Sun2025,li2025}. As the z$\sim$1 relation benchmark, the line of $\log(M_{\bullet}/M_*)=-2.5$ (black solid line) is plotted as well. It shows that there is little evidence for evolution up to z $\sim$ 4. The observed $M_{\bullet}$/$M_{*}$ ratio (before correcting for selection biases) beyond z $\sim$ 4 appears to increase with redshift, but we will show that this can be due to the selection biases.}
\label{Fig:delta_ratio_BH_z}
\end{figure*}

Figure~\ref{Fig:delta_ratio_BH_z} shows $M_\bullet/M_{*}$ vs.  redshift from z$\sim0$ to $7$, just for massive host galaxies, ($\log(M_{*}/M_\odot) \gtrsim 10$). There is a long history of controversy regarding whether it differs systematically at high redshift from the local relation. However, the figure shows that the data for high mass hosts, which have been the subject of virtually all the measurements for $0 < z < 4$, is consistent with little if any evolution for 0.1 $\lesssim$ z $<$ 4. Although it appears from Figure~\ref{Fig:delta_ratio_BH_z} that there is significant evolution for massive host systems for z $>$ 4, we will show that the offset can be explained in terms of selection effects, and a standard ratio may hold up to z $\sim$ 7.

There are other consequences of switching to the GSH20 foundation. The order of magnitude offset between the local RV15 sample and the values of $log(M_\bullet/M_{*})$ for samples at $z \gtrsim 0.1$  (see Figure~\ref{Fig:delta_ratio_BH_z}) is typically not treated quantitatively in previous works. However, to first order use of the GSH20 scaling relation accounts for this effect. As an example, for the $\log(M_\bullet/M_{\odot}) > 8.5$ sample of \citet{Suh2020}, the average {\it observed} offset from the RV15 relation is 1.03 in the log. In comparison, the {\it predicted} offset from the GSH20 ``all'' relation is 1.10, i.e., in good agreement. This result indicates how any high redshift sample can be compared via the GSH20 relation directly with local observations to test for evolution. To first order, the result is no evolution.

\section{Evolution of $M_\bullet/M_{*}$ with redshift}\label{sec:z_evol}

  We now discuss in detail  the redshift evolution of $M_{\bullet}-M_{*}$ relation up to z $\sim 7$ by including the two high-redshift comparison samples introduced in Section~\ref{sec:comparison_sample}, i.e., the z $\sim$ 6 Seyfert and quasar samples.
 From the above discussion, there are two areas that may affect  high $M_{\bullet}/M_{*}$ for the high redshift Seyferts: (1) the estimates of BH masses and host galaxy masses; and (2) selection biases such as the ``Lauer bias''\footnote{\citet{Lauer2007} pointed out that any study selecting AGNs on the basis of their luminosity will preferentially include ones with relatively massive SMBHs compared with a truly unbiased sample.}. We begin by a detailed disccussion of the mass estimates for both samples, and follow in Section \ref{sec:obs_bias_simu} with an analysis of the Lauer bias. 

\subsection{Unification of Seyfert $M_{\bullet}$ and $M_{*}$ measurements}
\label{unification}

The potential uncertainties  in black hole and galaxy masses are large, but it is nonetheless useful to put both on a consistent scale. \citet{Harikane2023} and \citet{Maiolino2023} used slightly different virial BH mass relations to infer BH mass for their  samples: \citet{Harikane2023} applied the original \citet{Greene2005} relation, while \citet{Maiolino2023} used one that incorporates the updated radius-luminosity relationship from \citet{Bentz2013}. Therefore, we recomputed BH masses for the \citet{Harikane2023} sample using the updated \citet{Greene2005}  relation to make it consistent with the \citet{Maiolino2023} sample.

Determining host galaxy masses for these quasars is challenging.  \citet{Harikane2023} determined masses by subtracting a PSF and fitting the residuals with \texttt{Prospector}. This approach was successful for only about half of their sample, since the residuals were inadequate for fitting the rest, {\bf which they show as upper limits}. 

The problem is more challenging at the order-of-magnitude lower masses of the \citet{Maiolino2023} sample.   As an example, we use the relation in  \citet{Allen2024} for z = 5 - 6 to estimate the expected effective radii for galaxies of the masses listed by \citet{Maiolino2023}; we get values of  0\farcs07 - 0\farcs1. The rest optical wavelengths, i.e. $\sim$ 4 $\mu$m observed,  are critical for mass estimation; but at 4 $\mu$m, the half-width at half maximum of the diffraction core of JWST is 0\farcs07. Therefore, for these low-mass cases, we do not expect useful residuals from PSF subtraction and have to estimate host galaxy masses from SED fitting. In \citet{Sun2025}, we compared masses estimated for AGN host galaxies with a \texttt{Prospector} - based SED fitting routine and found them to agree well for cases where we could meaningfully subtract a PSF and fit the residual, so we expect the results to be reasonably accurate.   

\citet{Maiolino2023} did SED-fitting with \texttt{Beagle} constrained by both NIRCam photometry (0.7 - 5 $\mu$m) and low-resolution (prism) NIRSpec data. They approximated the AGN SED as a reddened power law. Our fitting used \texttt{Prospector}, custom modified to provide accurate AGN SEDs \citep{Lyu2024}. It did not utilize a spectrum, but was constrained by 14 - 16 photometric bands (depending on the source), combining HST and JWST-JADES publicly released  photometry \citep{Rieke2023} and extending from 0.435 to 5 $\mu$m. These two approaches should return similar results for galaxy masses \citep{Pacifici2023} for the same initial conditions. However, the assumed star formation histories are different: (1)  \citet{Maiolino2023} assumed  a delayed-$\tau$ star formation history, with a burst of constant star formation 
lasting 10 Myrs prior to observation; and (2) we assumed a delayed-tau star formation history with no imposed tendency toward young stellar populations.  Our approach is similar to that of \citet{Harikane2023} and allows us to put all the galaxies on a similar basis. With some scatter, our mass estimates are greater than those of \citet{Maiolino2023} by about 0.5 dex. This difference likely arises from their assumption of a very young stellar population. It results in  $M_{\bullet}/M_{*}$ in the systems in their study tending to fall about 0.5 dex further from the zero-evolution case than do the same systems for us. The nature of this difference is illustrated in \citet[][Figure 5]{Stone2024}. 

We emphasize that the differing mass estimates are both ``correct'' mathematically. However, it is not necessarily appropriate to assume that {\it every} AGN lies in a host undergoing a starburst (i.e. vigorous star formation in the past 10 Myr). Although the nature of the star formation in AGN hosts remains controversial, the majority opinion is that they lie in galaxies on or below the main sequence \citep[e.g.,][]{Xu2015, Zhang2016, Schulze2019,  Vietri2022}. Therefore,  our approach based on the \texttt{Prospector} fits with no bias toward recent star formation is more conservative in the current context (shows less deviation of $M_\bullet/M_{*}$ from standard values). Our estimates also describe more accurately the majority of field galaxies at similar redshift  \citep[][Figure 5]{Stone2024}.

The unified parameters are listed in Table~\ref{tbl:unified}.

\begin{table*}[ht]
    \centering
    \begin{tabular}{cccc}
    \hline
    \hline
        Galaxy & z & $\log(M_{*}/M_{\odot})$ & $\log(M_{\bullet}/M_{\odot})$   \\
         \hline
  
    CEERS\_01244	&	4.478	&	${8.63}^{+1.03}_{-0.63}$	&	7.73	\\
GLASS\_150029	&	4.583	&	$9.1_{-0.31}^{+0.37}$	&	6.88	\\
CEERS\_00746	&	5.624	&	9.11	&	7.91	\\
CEERS\_1665	&	4.483	&	$9.92_{-0.51}^{+0.68}$	&		7.51	\\
CEERS\_00672	&	5.666	&	9.01	&	7.9	\\
CEERS\_02782	&	5.241	&	9.35	&		7.85	\\
CEERS\_00397	&	6	&	$9.36_{-0.36}^{+0.45}$	&		7.29	\\
CEERS\_00717	&	6.936	&	$9.61_{-0.77}^{+1.18}$	&	8.28	\\
JADES\_8083	&	4.648	&	$9_{-0.3}^{+0.3}$	&	7.25	\\
JADES\_1093	&	5.595	&	$7.98_{-0.3}^{+0.3}$	&	7.28	\\
JADES\_3608	&	5.269	&	$8.61_{-0.3}^{+0.3}$	&		6.91	\\
JADES\_11836	&	4.409	&	$9.03_{-0.3}^{+0.3}$	&	7.13	\\
JADES\_20621	&	4.681	&	$8.55_{-0.3}^{+0.3}$	&	7.34	\\
JADES\_77652	&	5.229	&	$8.63_{-0.3}^{+0.3}$	&	7.01	\\
JADES\_61888	&	5.875	&	$8.65_{-0.3}^{+0.3}$	&	7.23	\\
JADES\_62309	&	5.172	&	$8.74_{-0.3}^{+0.3}$	&	6.54	\\
JADES\_954	&	6.76	&	$9.34_{-0.3}^{+0.3}$&	7.9	\\

    \hline
    \end{tabular}
    \caption{Unified BH and stellar mass of the high redshift Seyfert galaxies (see Section~\ref{unification}) }
    \label{tbl:unified}
\end{table*}

\subsection{Quasar Host Masses}
\label{qsohostmasses}

\citet{Li2024} argue that the undetected host galaxies for high redshift quasars might have been missed because they are too compact and that their masses could easily exceed the upper limits that are derived from an assumption of significant extent for direct detection. This would undermine the usefulness of our quasar sample to analyze the behavior of $M_\bullet$/$M_{*}$ at the highest black hole masses. 
For our primary sample of 12 quasars, seven have host galaxy detections, so exploring the issue raised by \citet{Li2024} must be focused on the remaining five. For them, the upper limits are generally similar to the masses of the detected galaxies. There is no tendency for more compact galaxies at z $\sim$ 6 to be more massive than the more extended ones of similar mass \citep{Morishita2024, Varadaraj2024}, so it would be unlikely for {\it all} the most massive galaxies to be also the most compact. 

A variety of individual measurements support the validity of the assigned masses for these  high redshift quasar hosts, including those with only upper limits. For example, when they are available, dynamical masses leave little room for a massive but unseen compact galaxy. Simulations by \citet{Lupi2019} indicate that gas-based estimates underestimate the stellar velocity dispersion, and in general the dynamical masses based on gaseous emission lines are lower than our masses and upper limits from photometry, as expected. In addition, spectra centered on the quasar, in some cases, do not show the stellar features that would be expected from a massive but compact host galaxy.

Because of possible concerns about the host galaxy mass estimates, we discuss the individual objects below:

\begin{itemize}

\item{{\it J0100+2802}: The upper mass limit from PSF subtraction is $3.8 \times 10^{11}$ M$_\odot$ \citep{Yue2024}. From modeling ALMA measurements of the [CII] 158$\mu$m line, \citet{Neeleman2021} estimate a dynamical mass within a radius of $\sim$ 4 kpc of $1.02^{+0.61}_{-0.71} \times 10^{11}$ M$_\odot$. \citet{Wang2019} provide  an estimate of $1.9 \times 10^{11}$ M$_\odot$ within the same radius, based on the [CII] line originating in a rotating disk at the measured inclination; they emphasize that the systematic uncertainties could be large.  In this case, the upper limit from photometry is consistent with the expected underestimate from gas dynamics.} 

\item{{\it J0148+0600}: The host galaxy, of mass $\sim 4 \times 10^{10}$ M$_\odot$, is very well resolved by \citet{Yue2024}.}

\item{{\it J0731+4459}: \citet{Stone2024} use the lack of stellar features in a spectrum centered on the quasar to derive an upper limit to any very compact stellar population consistent with the limit they obtained for a more extended component from photometry, $ 5 \times 10^{10}$ M$_\odot$. }

\item{{\it J1030+0524}: There is no confirmation of the mass estimate of $< 4.5\times 10^{10}$ M$_\odot$ \citep{Yue2024}.}

\item{{\it J1120+0641}: A detection of the host has been reported by \citet{Marshall2024}. It appears to have a complex morphology, probably because it is a merging system. They derive a mass for the quasar host of $2.6^{+2.4}_{-1.4} \times 10^9$ M$_\odot$ from SED fitting. However, they have a detection only at 1.1 $\mu$m, in the rest far UV, making the SED fitting very uncertain as indicated by the large error. \citet{Stone2024} quote a measurement of the host at 3.6 $\mu$m of 0.20 $\pm$ 0.07 $\mu$Jy, while the detection in this band from \citet{Marshall2024} is 0.17 $\pm$ 0.08 $\mu$Jy. \citet{Stone2024} quoted an upper limit for the mass because of possible systematic errors if the host had a small diameter, but the image in \citet{Marshall2024} removes this concern and we can obtain a best estimate of 0.19 $\pm$ 0.05 $\mu$Jy for the flux density at 3.6 $\mu$m by averaging these measurements. Since there is insufficient information for a meaningful SED fit, we use the method of \citet{Stone2024} to find a host galaxy mass of $3.8 \pm 1.0 \times 10^9$ M$_\odot$. }

\item{{\it J1148+5251}: The upper limit to the galaxy mass from photometry is $< 5 \times 10^{10}$ M$_\odot$ \citep{Stone2024}, larger by roughly the expected amount \citep{Lupi2019} than the dynamical mass of $1.8 \times 10^{10}$ M$_\odot$ \citep{Willott2015}.}

\item{{\it J1340+2813}: \citet{Stone2024} use the lack of stellar features in a spectrum centered on the quasar to derive an upper limit to any very compact stellar population consistent with the limit they obtained for a more extended component from photometry, $< 6 \times 10^{10}$ M$_\odot$.} 

\item{{\it J159-02}: The host galaxy, of mass $\sim 1.4 \times 10^{10}$ M$_\odot$, is very well resolved by \citet{Yue2024}.}

\item{{\it J1512+4422}: The host galaxy is resolved by \citet{Onoue2024}. A  mass of $4.3 \times 10^{10}$ was determined from modeling the photometry. The H$\alpha$ and H$\beta$ lines are double-peaked, making a single-epoch black hole mass estimate challenging. We adopt the average from these two lines from \citet{Onoue2024}, $1 \times 10^{9}$ M$_\odot$.
}

\item{{\it J2236+0032}: The host galaxy is well resolved by \citet{Ding2023} who estimate  a mass of $13 \times 10^{10}$ M$_\odot$ from fitting the photometry. This mass is in reasonably good agreement with that derived by \citet{Onoue2024}, $6.5 \times 10^{10}
$. We have adopted the value from \citet{Onoue2024}.}

\item{{\it J2239+0207}: The host galaxy, of mass $\sim 1 \times 10^{10}$ M$_\odot$,  is resolved by \citet{Stone2023}. \citet{Stone2024} show that this mass is consistent with the lack of stellar features in the spectrum of the system. The dynamical mass is much larger, $\sim 29 \times 10^{10}$ \citep{Izumi2019}.}

\item{{\it J2255+0251}: The host galaxy, of $3.4 \times 10^{10}$ M$_\odot$, is well resolved by \citet{Ding2023}. }

\end{itemize}

 \citet{Marshall2023} have studied two additional high-redshift quasars using JWST integral field spectroscopy. 

\begin{itemize}

\item{ {\it DELS J0411-0907}:  The host galaxy mass of $8.6 \times 10^{10}$ M$_\odot$ was determined from dynamics based on optical emission lines. }

\item{{\it VDES J0020-3653}: The host galaxy mass of $17 \times 10^{10}$ M$_\odot$ was determined from dynamics based on optical emission lines. }

\end{itemize}

The method used for these latter two targets is seldom used on local galaxies and is not well calibrated relative to using photometry \citep{KH2013}. Because of possible biases in the galaxy mass determinations, we have not included these two galaxies in our analysis. However, their nominal positions in the $M_\bullet$ vs $M_{*}$ plane indicate that including them would not affect our conclusions. 

There are dynamical mass measurements from mm-wave line observations for a number of additional quasars. These are less useful for the current study because: (1) the $M_\bullet - M_{*}$ relation is defined for stellar galaxy masses and the dynamical masses may include significant contributions from the gas; and (2) the dynamical measurements are uncertain for a number of reasons, particularly the correction for inclination and the assumption that the source is a simple inclined disk.

There are also JWST discoveries of massive black holes at redshifts $>$ 7.5, but not yet within a systematic framework that would allow us to analyze their implications for the $M_\bullet$ - $M_{*}$ relation, \citep[e.g.,][]{Kokorev2023, Kovacs2024, Natarajan2024}. \citet{Rinaldi2024} report resolved images of three Little Red Dots (LRDs) with broad H$\alpha$ that could yield accurate host galaxy masses\footnote{although their complex morphologies are an obstacle}, but again the context is not yet well enough understood for a quantitative analysis.  Further discoveries of this nature along with additional host galaxy measurements for quasars at z $\sim$ 6 will substantially enhance our understanding of the $M_{\bullet}$ - $M_{*}$ relation at very high redshift in the future.

\subsection{Observational biases}\label{sec:obs_bias_simu}

After unifying stellar mass and BH mass measurements for the z $\sim$ 6 Seyfert sample, we remeasured their $M_{\bullet}$/$M_{*}$ ratios. The z $\sim$ 6 Seyferts and quasars still have an overall higher $M_{\bullet}$/$M_{*}$ ratio than expected from the scaling relation, leading to the suggestion that the SMBHs at $z>4$ are ``overmassive'' related to their host galaxies compared to the low redshift Universe.

However, observational biases can significantly reshape the observed quantities even if the true relation is the same (i.e., \citealt{Lauer2007,Li2024}), contributing to the apparent  
 difference in $M_{\bullet}/M_{*}$ between low and high redshift. Therefore, in this section, we apply a Monte Carlo simulation to explore the probability that the observed $z\sim6$ Seyfert and quasars still follow the local relation when considering the observational biases.

Our modeling approach to test for selection biases is described in detail in \citet{Sun2025}. Basically, we generate a mock AGN sample that represents the underlying AGN population at a specific redshift epoch and,  following the baseline relation, apply the observational effects on it to make a mock AGN sample, then test how likely it is that the observed distribution on the $M_{\bullet}$-$M_{*}$ diagram is consistent with the mock distribution. We start with an ideal scenario in which the survey area is infinitely large to explore the intrinsic  distributions on the $M_{\bullet}$-$M_{*}$ plane (Section~\ref{sec:inf}). We then simulate the actual observations with an appropriate number of mock AGNs (Section~\ref{sec:limited}). The simulations depend on a number of inputs: the stellar mass function (SMF), the Eddington Ratio Distribution Function (ERDF), and the broad line AGN fraction. We postpone discussion of the effects of uncertainties in these parameters themselves to Section 5.


\begin{figure*}
\centering
    \includegraphics[width=1\textwidth]{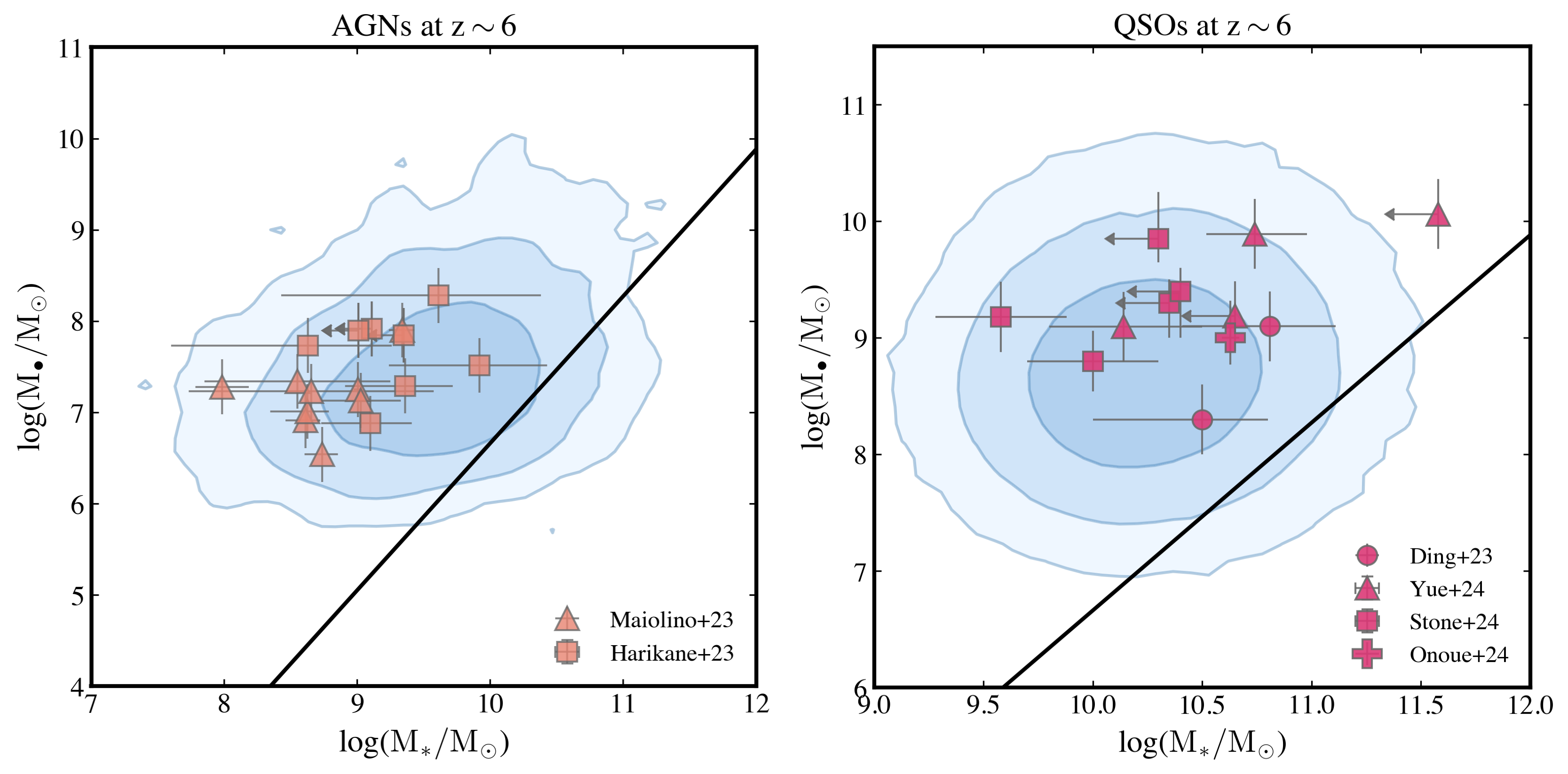}
\caption{The distributions on the $M_{\bullet}$-$M_{*}$ diagram at $4<z<7$ of the AGN sample (left) and the quasar sample (right), assuming the GSH20 relation with a scatter of 0.8 dex. 
The contours show the predicted distribution from our Monte Carlo simulation that includes both errors and biases, assuming an infinite number of tries. Specifically, the contours fall slightly above the input log($M_\bullet/M_{*}$) primarily because of the ``Lauer bias''. This result demonstrates that our simulation is capable of producing values similar to those observed. {\bf However, as shown in the following figure,  when we restrict the number of trials from virtually infinite to realistic values, we cannot account for the observations under the assumptions.}  
}
\label{Fig:observable_z6}
\end{figure*}


\begin{figure*}
\centering
    \includegraphics[width=1\textwidth]{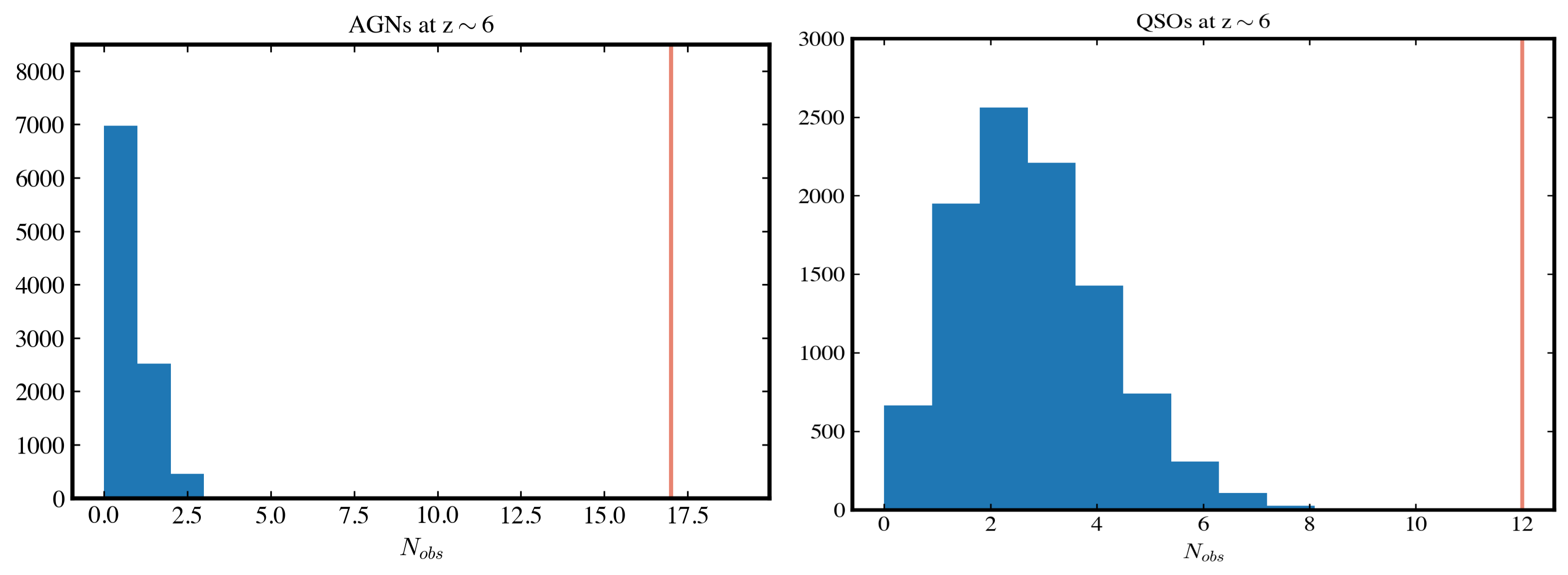}
\caption{The MC mock ``observable'' AGN number count distribution (blue histogram) of the z $\sim$ 6 AGN sample (left), and the z $\sim$ 6 quasar sample (right), assuming the underlying intrinsic relation is the GSH20 relation. The orange lines represent the observed number counts of the z $\sim$ 6 Seyferts ($N_{obs}=17$) and quasars ($N_{obs}=12$). That is, when we restrict the number of trials to match the observations, the simulations come short of the number of high $M_\bullet/M_{*}$ galaxies observed.}
\label{Fig:MC_all}
\end{figure*}


\subsubsection{Infinitely-large survey}
\label{sec:inf}

{\bf Seyferts:} For the Seyferts, we use $10^7$ random trials to draw $10^7$ mock BL AGNs with true stellar mass $M_{*\text{, true}}$ and redshift $z_{\text{true}}$ from the galaxy stellar mass function (SMF, \citealt{Weibel2024}). Assuming that the BL AGNs follow the GSH20 relation (with a scatter of 0.8 dex), we determine the true BH mass $M_{\bullet, true}$. We then adopt the intrinsic Eddington ratio distribution function (ERDF) at $z \sim 6$ from \citet{Wu2022} to determine the bolometric luminosities ($L_{bol}$). 
Next, assuming the virial BH mass and stellar mass measurements are not biased from the true mass but are subject only to measurement errors, we randomly add a Gaussian error with a dispersion of 0.3 dex to the $M_{\bullet\text{, true}}$, and 0.3 dex to the $M_{*\text{, true}}$ to obtain $M_{\bullet\text{, obs}}$ and $M_{*\text{, obs}}$. These errors were determined by \citet{Maiolino2023} and \citet{Harikane2023}.

Finally, we apply the observation limits to select the mock AGNs with properties similar to the observed samples. We note that the JADES, CEERS, and GLASS NIRSpec samples are not selected as flux-complete samples, resulting in complicated selection biases. Therefore, we assume a simplified scenario for which the selection biases of the Seyfert sample only include a broad $\mathrm{H\alpha}$ line width threshold for BL AGN identification ($\text{FWHM}_{\mathrm{H\alpha}}>1000\,{\rm km\,s^{-1}}$) and a spectral sensitivity limit. To estimate the FWHM of the broad line for mock AGNs, we applied the $\mathrm{H\alpha}$-based virial BH mass relation (i.e., the updated \citet{Greene2005} relation). To model the spectral sensitivity limit, given that JADES and CEERS NIRSpec data can typically reach a line flux sensitivity of $\sim5.5\times10^{-19}\,\mathrm{erg\,s^{-1}\,cm^{-2}}$ for 3$\sigma$ detection of a broad line of 1000$\,\mathrm{km\,s^{-1}}$  \citep[see ][]{Pacucci2023}, we convert this line flux sensitivity to the spectroscopic AGN bolometric luminosity limit $L_{\text{lim, spec}}$ to determine an AGN bolometric luminosity cut. 
Considering the two observation limits mentioned above, we made a mock AGN subsample from the whole AGN sample.

The left panel of Figure~\ref{Fig:observable_z6} shows that the $M_{\bullet\text{, obs}}$-$M_{*\text{, obs}}$ distribution of the mock z $\sim$ 6 Seyferts is strongly biased upwards from the intrinsic GSH20 relation, confirming the significant effects of the Lauer bias. The observation limits that only allow us to detect bright AGNs will bias the BH mass distribution to the massive end at a fixed $M_{*}$, which becomes stronger at the lower $M_{*}$ regime. More importantly, the mock distribution encloses the observed z $\sim$ 6 Seyferts well within the 3$\sigma$ distribution. This simulation shows that sources matching the observed  $M_{\bullet}-M_{*}$ relation at 4$<$z$<$7 can be produced by the simulation with a sufficiently large number of trials.

{\bf Quasars:} The same routine is applied to model the infinitely large survey of z $\sim$ 6 quasars. We assume the same z $\sim$ 6 SMF and ERDF for the z $\sim$ 6 quasars as  for the z $\sim$ 6 Seyferts, as well as the $M_{\bullet}$ and $M_{*}$ measurement errors (both are 0.3 dex). 
Similarly, the selection biases of the z $\sim$ 6 quasar sample also contain an AGN luminosity limit ($L_{bol}>10^{46}\,{\rm erg\,s^{-1}}$) and a cut of BL AGN line width ($\text{FWHM}_{\mathrm{H\beta}}>1000\,{\rm km\,s^{-1}}$). Taking into account those two observational limits, we find the distribution of the mock quasars at z $\sim$ 6 on the $M_{\bullet}$-$M_{{*}}$ plane as shown in
the right panel of Figure~\ref{Fig:observable_z6}. Again, we find that the mock z $\sim$ 6 quasar distributions are significantly shifted above the original GSH20 relation. Also, most of the observed z $\sim$ 6 quasars are located within the 2$\sigma$ contour, although {\it J0100+2802} is outside of the mock distribution. Given that the stellar mass measurement of {\it J0100+2802} is only an upper limit and it could lie within the mock distribution if its real mass were 0.2 dex lower, we conclude that our mock AGN simulation can successfully reproduce the observed z $\sim$ 6 behavior given a sufficiently large number of trials. 

For both the Seyferts and Quasars, our results are in agreement with those of \citet{Li2024}.

\subsubsection{Area-limited survey}
\label{sec:limited}

While Figure~\ref{Fig:observable_z6} shows that the $M_{\bullet}$-$M_{{*}}$ distribution of the mock BL AGNs of an infinitely large survey is able to enclose the truly observed z $\sim$ 6 Seyferts and quasars, it does not necessarily indicate that these data points are consistent with the mock distribution. With a large enough sample, the distribution can successfully include the outliers, which may just reproduce some observed targets by chance. Therefore, as we introduced in \citet{Sun2025} in detail, we make an apples-to-apples comparison, in which the mock AGNs are modeled as we would find from a set of observations (i.e., with a comparable sample size and parameter ranges ($M_{*}$ and $z$)). The consistency between mock and true $M_{\bullet}$/$M_{{*}}$ distribution can be investigated using the K-S test. We find, as discussed below and illustrated in Figure ~\ref{Fig:MC_all}, that the area-limited simulation matches the actual observations poorly, especially for the Seyfert case.  

{\bf Seyferts:} Given a parent sample size of the JADES and CEERS NIRSpec surveys of 758 (Ignas Juodzbalis, private communication) and 185 \citep{Harikane2023} at $4<z<7$, respectively, we directly simulate this number of galaxies over the stellar mass range of $7.8<\log {M_{*}/M_{\odot}}<10.7$ using the same SMF that we used for the infinitely-large survey. Namely, we simply assume that the NIRSpec target selection for these 943 galaxies does not have a stellar mass bias that deviates from the SMF.
Next, we run this number of trials to simulate the mock galaxy population assuming the GSH20 relation. 
We then apply a BL AGN fraction (the number ratio of BL AGNs to galaxies) from \citet{Maiolino2023} ($10$\%) to this population to estimate the number of galaxies (or trials) to have a BL AGN (i.e. $\sim100$ BL AGNs).
After assigning an Eddington ratio, mass measurement uncertainties, and observational limits, as we did in Section~\ref{sec:inf}, we generate a mock observation for the Seyfert sample.

By repeating the procedure of mock observation generation 10000 times, we determine the distribution of the ``observable'' Seyfert  number counts expected in each mock observation with the intrinsic ratio determined by the GSH20 relation. This result is plotted in the 
left panel of Figure~\ref{Fig:MC_all}. The distribution indicates that, in most of the MC iterations, the predicted number count of ``observable" Seyferts  at z $\sim$ 6 within the Harikane and Pacucci samples is zero and only a few runs can produce 1--2 ``observable" Seyferts, which is much smaller than the observed number count ($N_{obs}=17$). 

Our result disagrees with the conclusion of \citet{Li2024}, who concluded that selection biases could account for the apparently overmassive black holes in \citet{Harikane2023, Pacucci2023}. Our run for an infinitely large survey shown in Figure~\ref{Fig:observable_z6} agrees with their results and shows that our modeling framework is consistent with theirs. However, in our area-limited survey simulation, we have reproduced the number of cases actually included in the studies and assessed the correspondence with our models through a probability calculation. In comparison,  they have estimated the potential number of cases by assuming the NIRspec target is a complete flux-limited sample drawn from a definitive survey area, which is uncertain given the complicated NIRSpec target selection functions in JADES and CEERS.
It is important to simulate the correct number of observations, and doing so shows that biases alone cannot account for the observed number of Seyferts lying above the scaling relation. Moreover, their area-limited simulation (see Figure 4 in \citet{Li2024}) showed that either the  \citet{KH2013} or GSH20 scaling relation could reproduce the observed BL AGN number count, but only if the scatter was increased for  GSH20.  In comparison, we have found (Figure~\ref{mbh_mstars}) that GSH20 is strongly preferred (compare the fit to the data, ``AGN total
''  with ``Greene, all'' and then witrh ``KH13''  in the figure). Our selection of the appropriate scaling relation contributes to the evidence that biases cannot fully explain the overmassive AGN, as we have shown in Section~\ref{sec:baseline}.

Although our unification of the measurements and use of the GSH20 scaling relation in place of the RV15 one have reduced the size of the effect, the mismatch of the ``observable'' Seyfert number count between our simulation and the observations demonstrates that the conclusions of \citet{Harikane2023, Pacucci2023} still hold. The observed incidence of massive black holes exceeds the predictions of the GSH20  relation extrapolated from $\log(M_{*}/M_{\odot})$ = 10  down to the relevant masses, i.e., $log(M_{*}) \sim$ 8.5 - 9.5. However, Figure~\ref{mvsm} indicates that some or all of the discrepancy may stem from the GSH20 scaling relation not holding at the lower masses, rather than to cosmic evolution in the $M_\bullet/M_{*}$ ratio for z $\ge$ 4, a possibility we will return to in Section~\ref{sec:discuss}. 

Also, due to the lack of predicted ``observable'' Seyferts, it is not possible to  make a further comparison between the $M_{\bullet}$/$M_{{*}}$ distributions of the mock observations and the true one using the K-S test. However, we will do the distribution comparisons in Section~\ref{sec:discuss} when we will test other  possibilities to match the mock Seyfert number count to the observed value.

{\bf Quasars:} We now make the mock observation simulation for the z $\sim$ 6 Quasars. The high-z quasars studied by \citet{Ding2023, Stone2024, Yue2024, Marshall2024, Onoue2024} do not come from an area-limited survey but are arbitrarily selected from multiple surveys to cover specific parameter spaces. In this case, we roughly estimate the minimum galaxy population size required for observing 12 luminous quasars with $L_{bol}>10^{46}\, {\rm erg\,s^{-1}}$ from the quasar luminosity function (QLF) at z $\sim$ 6 reported by \citet{Shen2020}. We find the survey area would  need  to be  $\sim3\times10^{4}\,{\rm arcmin^2}$. This area contains $1.4\times 10^3$ galaxies with $10<\log (M_*/M_{\odot})<11.6$ at 5$<$z$<$7.1. Again, we apply the same BL AGN fraction as was assumed for the z $\sim$ 6 Seyferts and then incorporate the observational limits to generate a mock observation of z $\sim$ 6 quasars. The MC distribution of the mock quasar number counts at z $\sim$ 6 predicted based on the GSH20 relation is shown in the right panel of Figure~\ref{Fig:MC_all}. 
The predicted quasar number count is about 2, six times smaller than the  observed value ($N_{obs}=12$). As before, these small numbers derived from the original GSH20 relation do not allow us to pursue the $M_{\bullet}$/$M_{{*}}$ distribution comparison between each mock observation and the true observations.

\subsubsection{Dependence of simulation results on assumed scatter in $M_\bullet/M_{*}$}

Considering the very young age of the universe at $4<z<7$ ($<1\,\text{Gyr}$), it is plausible that the intrinsic scatter of the mass scaling relation at high-z is larger than it is at z $\sim$ 1.5, as indicated by simulations  \citep[e.g.,][]{Hirschmann2010, Huang2018}. Here we evaluate how large an increase in the assumed scatter could bring the results reported in the preceding section into agreement with the GSH20 scaling relation. 

{\bf Seyferts:} 
We first evaluate the situation for the Seyferts. We see whether plausibly larger values for the scatter could increase the median value of the MC number count distribution to be consistent with the observed value ($N_{obs}=17$),  keeping the slope and intercept the same as the GSH20 relation. 
 We find the scatter around the GSH20 relation has to be at least 4.7 dex, 6 times larger than the local scatter. 

\begin{figure*}
\centering
    \includegraphics[width=1\textwidth]{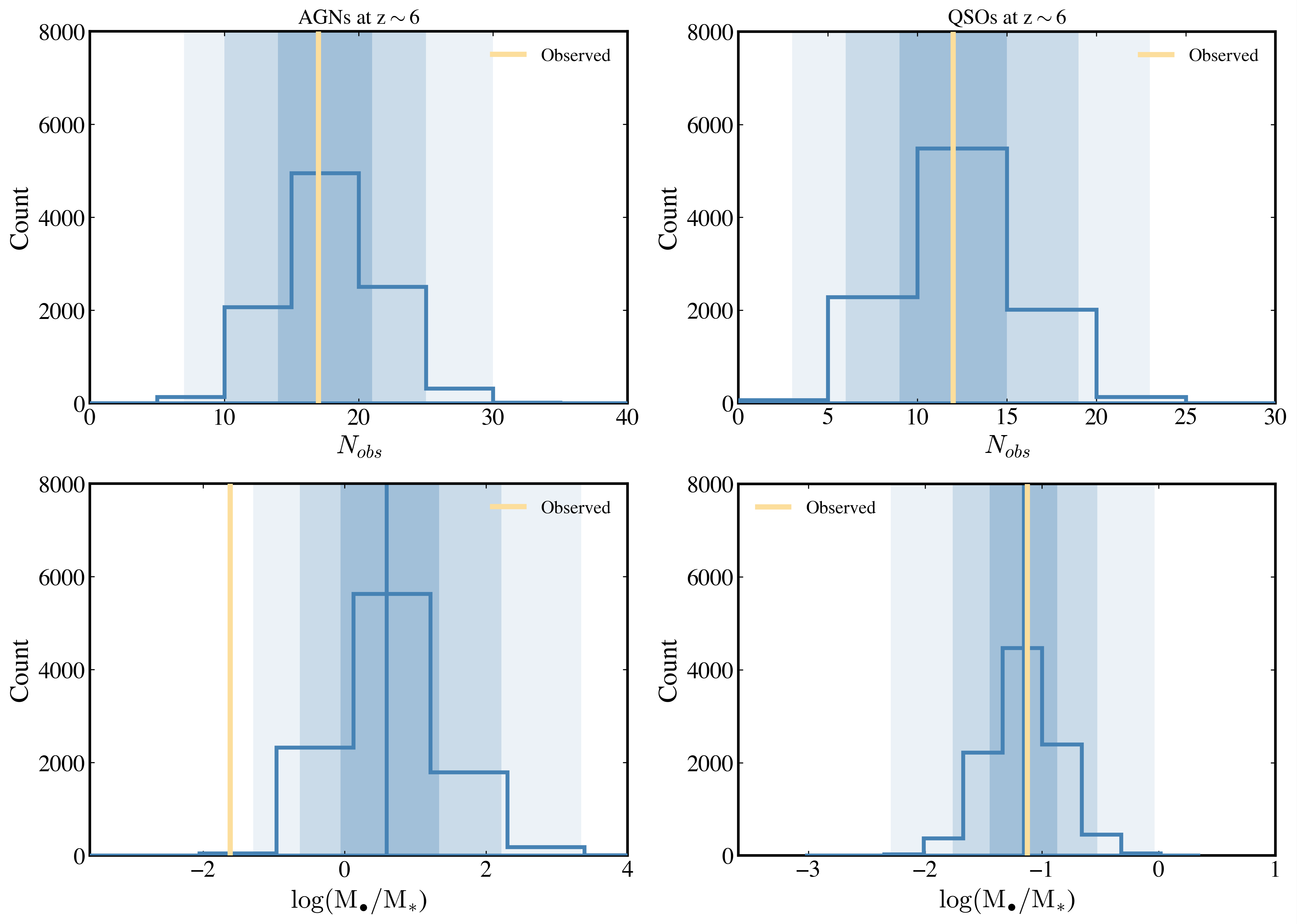}
\caption{MC simulation results for the mock z $\sim$ 6 Seyfert sample (left) and quasar sample (right) assuming the underlying intrinsic relation is the GSH20 relation with a larger scatter (4.7 dex for Seyferts and 1.2 dex for quasars) than that of the low-z relation (0.8 dex). The blue histograms represent the mock distributions and the yellow lines represent the observed value for the two samples.}
\label{Fig:MC_all1}
\end{figure*}

After matching the number count assuming this level of scatter, we then do a K-S test between the $M_{\bullet}$/$M_*$ distributions of each mock observation and the true observations to test the consistency; namely, we make 10000 K-S tests and calculate the fraction of test results for which we can reject the hypothesis that the observed AGN sample comes from the same distribution as the mock AGNs (the p-value of the K-S test is higher than 0.05, where values $<$0.05 would imply a significant difference). This approach accounts for the fluctuations that exist in our small sample of observations and provides a statistical probability of rejecting the null hypothesis. The K-S test result shows that, in $>99\%$ (9987/10000) of the realizations of the two-sample K-S test for the GSH20 relation with a larger scatter of 4.7 dex, the p-values are $<$ 0.05, indicating that the observed $\log (M_{\bullet}$/$M_{*})$ distribution is different from the mock distribution derived from this assumed baseline relation. Also, the median $M_{\bullet}$/$M_{*}$ of the mock observable sample produced with such a larger scatter is still $>3\sigma$ larger than the observed value (the left column of Figure~\ref{Fig:MC_all1}). Therefore, only increasing the scatter of the GSH20 relation still cannot fully reproduce the observed high $\log (M_{\bullet}$/$M_{*})$ ratios of the BL Seyferts at $4<z<7$ after taking the observational biases into account. In other words, the slope and intercept of the $M_{\bullet}$/$M_{*}$ relation at z $\sim$ 6 would also need to be different from the values of the local GSH20 relation.

{\bf Quasars:} Similarly to the z $\sim$ 6 Seyferts, we test how much the scatter around the  scaling relation has to increase to match the predicted z $\sim$ 6 quasar number count with the observed one. We find that an increase of the scatter of the GSH20 relation from 0.8 dex to 1.45 dex is needed. The modification of the underlying distributions required for matching the observed Quasar number counts is much less than for the Seyfert case. Again, we run the K-S test between the $M_\bullet$/$M_{*}$ distribution of each mock observation predicted by the GSH20 relation with a larger scatter and of the observed quasar sample and calculate the fraction of the MC realizations with the p-value of the K-S test higher than 0.05 (supporting the same distribution scenario). The MC result shows that most of the realizations (9799/10000, $>97\%$) have a p-value of $>0.05$, indicating the $M_\bullet$/$M_{*}$ distribution of the mock observations derived from the GSH20 relation with a scatter of 1.45 dex can be consistent with the observed distribution when taking the observational biases into account (the right column of Figure~\ref{Fig:MC_all1}).

\section{Discussion} \label{sec:discuss}

The simulations just reported imply that $M_\bullet/M_{*}$ is high for both samples. We now integrate our results with other information to evaluate whether  cosmic evolution of $M_\bullet/M_{*}$ is {\it required} to explain the observations at z $>$ 4, or whether the data could be consistent with this relation behaving roughly as is observed at lower redshift. 

Our simulations incorporate several important parameters regarding the properties of the whole galaxy and AGN populations at $4<z<7$ that might affect the results: the SMF, ERDF, and BL AGN fraction. Those basic distributions have not been fully constrained at high redshift, especially the ERDF and BL AGN fraction. Therefore, inappropriate assumptions for any of them that deviate from the true distribution at z $\sim$ 6 could cause a number count discrepancy between the mock observation and the true observation. 

\subsection{Seyferts}

In this section, we test the impact of those three parameters on the mock observation simulations for the Seyferts. We find no modifications that can significantly affect the high black hole masses compared with the extrapolated GSH20 scaling relation. Therefore, at the end of the section,  we consider alternatives to this baseline scaling relation.  

\subsubsection{Stellar mass function}

In the above simulations, we used the SMF from \citet{Weibel2024} derived using JWST deep imaging survey data (CEERS, JADES, and PRIMER). This SMF is built based on the stellar masses measured using data out to NIRCam F444W, giving coverage to the \textit{V} band at z $\sim$ 6 -- 7, and is well determined at the stellar mass range of $8<\log ({M_{*}/M_{\odot}})<10$. Therefore, it perfectly covers the range of the $z \sim 6$ Seyfert stellar masses, even though it has only upper limit constraints at the high stellar mass end. Also, within this stellar mass range, the \citet{Weibel2024} SMF is consistent with the other commonly-used SMFs, e.g.,  \citet{Weaver2023} and \citet{Wang2024}. Therefore, for the Seyfert case, the current SMF can produce a population with a proper number of galaxies with similar stellar masses to those observed for the high-z faint AGNs. In addition, as we mentioned in Section~\ref{sec:obs_bias_simu}, we simply assumed the parent sample that we are drawing from is unbiased from the whole galaxy population at $z\sim$ 6 and follows the SMF, which may not be correct given the complicated NIRSpec follow-up target selections for the JADES and CEERS surveys \citep[e.g.,][]{Bunker2024}. However, since the SMF at z $\sim$ 6 is already dominated by the low-mass galaxies ($\log({M_{*}/M_{\odot}})<10$), even if the parent sample is biased more towards the lower-stellar mass end, the corrected stellar mass distribution from the SMF cannot recover six times more mock ``observable'' BL AGNs in the simulation, i.e., plausible uncertainties in the SMF do not help solve the issue of underestimated BL AGN number count.

\subsubsection{Eddington ratio distribution function}

Therefore, we have tested the impact of the assumed ERDF on our simulation results by shifting the peak of the ERDF upwards to -0.5 (corresponding to the 1$\sigma$ error of the best-fit ERDF; see Figure 5b in \citealt{Wu2022}). We found that an ERDF weighted more at the higher end does not increase the predicted ``observable'' BL AGN number counts for the Seyfert case.  This is because the GSH20 relation infers an 
 extremely small $M_{\bullet}$ for mock galaxies with stellar mass ranging from $7.8<\log (M_{*}/M_{\odot})<10.7$. For an AGN with such a low $M_{\bullet}$,  a higher Eddington ratio boosts the AGN luminosity, and since the simulation is built around the black hole masses, the inferred broad line width using the H$\alpha$ (or H$\beta$)-based single-epoch BH mass relation will decrease at the same time. Consequently, the AGN will fail the broad line width cut in the simulation. Another issue can be seen from Figure~\ref{mvsm}. Higher Eddington ratios will result in overestimates of the black hole masses, but since the black hole masses are an order of magnitude above the GSH20 relation, there is simply not enough latitude in the assumed Eddington ratios to bring them down to the relation. 

\subsubsection{BL AGN fraction}

 Our simulation could predict more ``observable'' BL AGNs by just increasing the assumed BL AGN fraction. from the assumed 10\% \citet{Maiolino2023}. However, we find that, to even make the 3$\sigma$ upper limit of the MC number count distribution consistent with the observed value, we have to boost the BL AGN fraction up to 85\% for the Seyfert case which is much higher than the well-constrained BL AGN fraction in the lower redshift Universe ($z\leq2$). Such a high value of BL AGN fraction is also unlikely since many  of the high-$z$ AGNs are obscured \citep[e.g.,][]{Lyu2024}

\subsubsection{Modified scaling relation}
\label{modscale}

 In summary, uncertainties in none of the three underlying parameters can solve the number count mismatch relative to the GSH20 scaling relation for the z$\sim$6 Seyfert sample. However, Figure~\ref{mvsm} shows that the extrapolation of the GSH20 relation is probably {\it not} appropriate. Instead, there is a population of low mass galaxies at 0.5 $<$ z $<$ 3 that have $M_\bullet/M_{*}$ ratios very similar to those of the z $=$ 4 -- 7 Seyferts \citep{Mezcua2023, Mezcua2024}. These lower-redshift AGNs are drawn from a sample very similar in size to the parent sample for the z $=$ 4 -- 7 cases, and they show a comparable number of galaxies with similarly high values of $M_\bullet/M_{*}$. Hence, selection biases should be similar and not affect this result. Thus, the z = 4 - 7 Seyferts do not necessarily indicate a status that fades away quickly after their birth and hence is only seen at very high redshift. Instead, the existence of low mass galaxies with similar properties from z = 0.5 to 3 indicates that they may be members of a previously poorly recognized low-mass galaxy population with relatively massive central black holes. That is, as stated by \citet{Mezcua2024}, ``from a statistical
perspective ... the z $>$ 4 sample
discovered by JWST and the VIPERS sample at z $<$ 3
... belong to the same population.'' \citet{Zhang2024} reinforces this suggestion by showing that the two populations may lie on the same evolutionary tracks.

The existence of a large population of low mass galaxies ($\log(M_*/M_\odot) \le 9.5$) near cosmic noon and with relatively luminous AGNs has also been inferred from studies of the population of AGNs in the CEERS \citep{Yang2023} and SMILES \citep{Lyu2024} fields. \citet{Backhaus2023}  identify a similar population at $0.6 < z < 1.3$ from HST observations. As shown in Figure~\ref{massdist}, these galaxies are rare locally, but a variety of evidence is revealing a major AGN population in low mass galaxies at cosmic noon and higher redshifts. 

Where are the counterparts at z $\sim$ 0 of these low-mass galaxies with high $M_\bullet/M_{*}$ at cosmic noon? In this redshift range and at their masses, the growth of their black holes is likely to proceed significantly more slowly than their stellar masses \citep[][Figure 2]{Zhang2024}, \citep[also, e.g.,][]{Aird2010,Calhau2017, McAlpine2017}.  If we assume that they lie on or above the star-forming main sequence 
\footnote{
which already emerged even at the stellar mass regime below their masses \citep{McGough2017} ($\sim10^8\,M_{\odot}$) at z$\sim$2 \citep{Merida2023}.}, then their star formation rate densities \citep{Popesso2023} are sufficient for significant stellar mass growth. 
For example,  main sequence galaxies with masses $\ge$ 10$^9$ M$_\odot$ at $z\sim1$ would grow into masses of $\gtrsim$ 10$^{10}$ M$_\odot$
\footnote{It is likely that these galaxies have sufficient gas to support substantial star formation. Local main sequence galaxies in this stellar mass range are gas-rich, with sufficient gas to, on average, triple their masses \citep{McGough2017}. A significantly larger molecular gas reservoir is expected at $z\sim$ 1 -- 2, in part from accretion of gas \citep{Walter2020}.} 
over the $\sim$ 8 Gyr from $z\sim1$ to $z\sim0$. These two trends - the slowed SMBH growth and the main-sequence level stellar mass gain - indicate that the descendents of these galaxies may now be hiding in plain sight with current masses $\gtrsim$ 10$^{10}$ M$_\odot$ and ratios of $M_\bullet/M_{*}$ within the local scaling relation.  

This population needs to be included quantitatively in future studies of AGNs. It suggests that the growth of black holes may at times outstrip the growth of low-mass host galaxies relative to the Magorrian relation for $z \ge 0.5$, and that the galaxy mass ``catches up'' during periods of SMBH quiescence.

\subsection{Quasars}

The results of increasing the scatter around the local GSH20 relation indicate that a modestly larger scatter than locally can solve the issue of the underestimated ``observable"  number counts for the z $\sim$ 6  quasars. There are other possibilities to resolve the discrepancy in number counts also, discussed below. 

\subsubsection{Stellar Mass Function}

Contrary to the situation with the Seyferts, the differences among the estimated $z \sim 6$ SMFs at the high mass end are quite large. For example, the $z\sim6$ SMFs from \citet{Weaver2023} and \citet{Wang2024} in the range of $\log ({M_{*}/M_{\odot}})>10$ are approximately 1 dex higher than the \citet{Weibel2024}  SMF. Therefore, one may wonder whether the missing BL AGN problem is due to an  underestimation of the high-mass SMF in this stellar mass range by \citet{Weibel2024}.

There are large uncertainties at the high-mass end ($\log ({M_{*}/M_{\odot}})>10$) for all three SMFs, as a result of limited number counts, and different extrapolations from lower masses. 
For example, the SMFs derived from the JWST data (e.g., \citet{Weibel2024} and \citet{Wang2024}) only used $\sim$300--500 arcmin$^2$ of imaging data; even for the \citet{Weaver2023}'s SMF derived from the COSMOS survey with a bigger area ($\sim1.27$ deg$^2$), the number density points above $\log(M_*/M_\odot)>10^{10.5}$ still have much larger error bars compared to the lower mass end. All three studies use Schechter functions extrapolated into high galaxy mass ranges where the data are very noisy. The differences among the three SMFs actually come from the fitted Schechter functions rather than the observed number densities, which may be due to the different constraints on the parameters.

As a measure of the resulting uncertainties, if we adopt the \citet{Weaver2023} or \citet{Wang2024} SMF, it is easier to match the predicted ``observable" quasar number count to the observed number count with a much smaller increase of the intrinsic scatter of the GSH20 relation (from 0.8 dex to 0.9 dex).

\subsubsection{Eddington ratio distribution function}
This issue does not occur in the quasar simulation because the GSH20 relation at higher stellar mass range can supply a high enough $M_{\bullet}$ ($\log(M_{\bullet}/M_{\odot})\sim7.5$) for an AGN to still have a sufficiently broad line to remain above our selection threshold.  
\subsubsection{BL AGN fraction}

Here the situation is not as extreme as for the Seyferts, but still the BL AGN fraction would need to be raised to  30\% to achieve consistency with the observed number counts. This would be difficult to reconcile with evidence that the UV duty cycle of UV-luminous quasars at z $>$ 6 is $\ll$ 1 \citet{Eilers2024}, see also \citet{He2018}.

\subsection{Outliers}

We have generally limited our analysis to sources at $z \lesssim 7$ since the conditions for detailed analysis of the systematics of the detections are not in place for higher redshifts. Nonetheless, there are instances where it appears that 
SMBHs at very high redshift can have masses comparable to those of their host galaxies, well outside the expected Magorrian ratios:

\begin{itemize}

\item{\citet{Kokorev2023} report an object at z = 8.50 with a black hole of $\log(M_\bullet/M_\odot) = 8.17 \pm 0.42$ and for which modeling of the SED indicates $\log(M_*/M_\odot) < 8.7$.}

\item{\citet{Natarajan2024} study a lensed galaxy at $z \approx 10.1$ with a black hole of $\log({M_\bullet/M_\odot}) \approx 7.6
$ and in a host galaxy of similar mass.}

\item{As shown in  Figure \ref{mvsm}, a number of LRDs appear to have black hole masses as large or nearly so to their host galaxy masses.}

\item{As discussed in Section \ref{qsohostmasses}, the host mass of the quasar J1120+0641 (z = 7.08) is only a factor of 2 - 3 greater than its SMBH mass of $\sim 1.4 \times 10^9$ M$_\odot$. 
This case will be discussed in depth in a forthcoming paper (Meredith Stone, in preparation).}

\end{itemize}
These examples indicate strongly that their black holes started to grow from massive seeds, rather than stellar mass ones  \citep{Natarajan2024,  Jeon2025}. They also imply that, our finding that black holes in massive galaxies cannot be shown to differ from the local scaling relation, is likely to break down at very high redshifts, and might even be shown to be fraying at $z \approx 6$ with better statistics than were available to us.

They also raise questions about whether these systems will ever evolve through mergers and star formation \citep[e.g.,][]{Rodriquez2016, Husko2023} to fall within the standard $M_\bullet/M_{*}$ relation. There may be a second evolutionary path  - for the host galaxies to remain low mass and the system to evolve into a virtually isolated SMBH. There are at least a few relevant examples \citep[e.g.,][]{Schramm2019, Liepold2025}. This may be consistent with the near-universality of the Magorrian relation \citep{KH2013}, since the detection of a massive black hole through stellar velocities or the presence of an AGN will be strongly biased toward examples only in major galaxies.

\section{Conclusion}\label{sec:conclusion}

We have carried out a comprehensive review and discussion of arguments for the evolution of $M_\bullet/M_{*}$ from high redshift to the current epoch. We found that the local scaling relation of GSH20 is preferred as a baseline reference for $\log(M_{*}/M_\odot) > 10$. Using this relation, we find that we can link local AGNs to higher redshift ones quantitatively assuming no change in the $M_\bullet/M_{*}$ behavior. However, there is no high-weight scaling relation to be used for $\log(M_{*}/M_\odot) < 9.5$ and the relation is based on small numbers between these ranges.

We confirm the existence at $z=$ 4 -- 7 of low-mass galaxies with relatively large central black hole masses,  significantly over-massive compared with an extrapolation of the GSH20 scaling relation.  However, there are also low-mass galaxies with similarly over-massive black holes down to $z \sim 0.5$ \citep{Mezcua2023,Mezcua2024}. 
  Thus, the case for evolution of $M_\bullet/M_{*}$ from z $\sim$ 6 to z $\sim$ 0.5 is not clear. However, low-mass galaxies with similarly high $M_\bullet/M_{*}$ are rare locally; the
  local galaxies with similar stellar mass
  have central black hole masses an order of magnitude lower than the ones found at z $\ge$  0.5 \citep[][Figure 2]{Greene2020}. That is, there does seem to be some form of evolution from z $\sim$ 0 to z $\sim$ 0.5. 

LRDs appear to have behavior similar to that of the classical 
AGNs in low mass hosts, i.e., $M_\bullet/M_{*}$ well above the standard relation for systems with massive hosts. Although we do not yet have a sample adequate for systematic analysis, this similarity argues for this behavior to be intrinsic, not the product of selection biases.

To account for the lack of local analogs, it is likely that the SMBHs in these galaxies have grown in a period of rapid accretion but this growth will slow, allowing the host galaxies to ``catch up'' and move the systems up in mass and closer to the standard scaling relation. 
 
The case of the quasars is simpler; the available sample is consistent with the GSH20 scaling relation, but at its extreme limits. These systems may  continue to evolve toward the present epoch retaining their atypically high $M_\bullet/M_{*}$ ratios, subject to the constraint of the availability of gas to accrete \citep[e.g.,][]{Zhang2024, Husko2023}. However, this picture becomes challenging for the most extreme cases such as ULASJ1120. For it, SMBH mass estimates of $1.35 \times 10^9$ M$_\odot$ \citep{Yang2021} or $1.4 \times 10^9$ M$_\odot$ \citep{Marshall2024} compare with our stellar mass estimate of $3.8\times 10^9$ M$_\odot$, i.e, 
$M_\bullet/M_{*}$ $\sim$ 0.3.  

It is possible that we are seeing, for both the low-mass systems and the quasars,  some cases where intense accretion can cause the SMBH to grow  beyond the standard $M_\bullet/M_{*}$ relation, but this episode is followed by a period of black hole quiescence.  Substantial galaxy growth by some combination of star formation and mergers then drives the system back to a more typical $M_\bullet/M_{*}$, a sequence that has also been proposed by, e.g., \citet{Tripodi2024,  Terrazas2024} and \citet{Bigwood2024}.

\section{Acknowledgements}

We thank Ignas Juodzbalis for providing the target count for the \citet{Pacucci2023} sample and George Mountrichas for sending us the data from his study. This work was supported by NASA grants NNX13AD82G
and 1255094. This work is based on
observations made with the NASA/ESA/CSA James Webb
Space Telescope. The data were obtained from the Mikulski
Archive for Space Telescopes at the Space Telescope Science
Institute, which is operated by the Association of Universities for Research in Astronomy, Inc., under NASA contract
NAS 5-03127 for JWST. These observations are associated with programs 1180 and 1181, for which the data can be accessed via \dataset[https://doi.org/10.17909/8tdj-8n28]{https://doi.org/10.17909/8tdj-8n28} \citep{https://doi.org/10.17909/8tdj-8n28}.

\facilities{JWST}

\bibliography{reference}{}
\bibliographystyle{aasjournal}



\end{document}